\documentclass[10pt,letterpaper]{article}

\usepackage[utf8]{inputenc}

\usepackage[super,sort&compress,comma]{natbib}
\usepackage{mhchem}
\DeclareRobustCommand{\citen}[1]{%
  \begingroup
    \romannumeral-`\x 
    \setcitestyle{numbers}%
    \cite{#1}%
  \endgroup   
}

\usepackage{nameref,hyperref}

\usepackage{microtype}
\DisableLigatures[f]{encoding = *, family = * }

\usepackage{changepage}

\usepackage[aboveskip=1pt,labelfont=bf,labelsep=period,singlelinecheck=off]{caption}

\makeatletter
\renewcommand{\@biblabel}[1]{\quad#1.}
\makeatother

\usepackage{lastpage,fancyhdr,graphicx}
\usepackage{epstopdf}
\fancyheadoffset[L]{2.25in}
\fancyfootoffset[L]{2.25in}

\usepackage{color}

\definecolor{Gray}{gray}{.25}

\usepackage{graphicx}

\usepackage{sidecap}

\usepackage{wrapfig}
\usepackage[pscoord]{eso-pic}
\usepackage[fulladjust]{marginnote}
\reversemarginpar

\begin{document}
\vspace*{0.35in}

\begin{flushleft}
{\Large
\textbf\newline{Transient Resonant Auger-Meitner Spectra of Photoexcited Thymine}
}
\newline
\\
Thomas J. A. Wolf,\textsuperscript{1,*},
Alexander C. Paul\textsuperscript{2}, 
Sarai D. Folkestad\textsuperscript{2},
Rolf H. Myhre\textsuperscript{2},
James P. Cryan\textsuperscript{1},
Nora Berrah\textsuperscript{3},
Phil H. Bucksbaum\textsuperscript{1,4},
Sonia Coriani\textsuperscript{2,5},
Giacomo Coslovich\textsuperscript{6},
Raimund Feifel\textsuperscript{7},
Todd J. Martinez\textsuperscript{1,8},
Stefan P. Moeller\textsuperscript{6},
Melanie Mucke\textsuperscript{9},
Razib Obaid\textsuperscript{3}, 
Oksana Plekan\textsuperscript{10},
Richard J. Squibb\textsuperscript{7}, 
Henrik Koch\textsuperscript{11*},
Markus G\"uhr\textsuperscript{12,*}
\\
\bigskip
\bf{1} Stanford PULSE Institute, SLAC National Accelerator Laboratory, 2575 Sand Hill Road, Menlo Park, CA 94025, USA.
\\
\bf{2} Department of Chemistry, Norwegian University of Science and Technology, NO-7491 Trondheim, Norway.
\\
\bf{3} Department of Physics, University of Connecticut Storrs, 2152 Hillside Road, Storrs, CT 06269, USA.
\\
\bf{4} Departments of Physics and Applied Physics, Stanford University, 382 Via Pueblo Mall, Stanford, CA 94305, USA.
\\
\bf{5} DTU Chemistry, Technical University of Denmark, DK-2800, Kongens Lyngby, Denmark.
\\
\bf{6} Linac Coherent Lightsource, SLAC National Accelerator Laboratory, 2575 Sand Hill Road, Menlo Park, CA 94025, USA.
\\
\bf{7} Department of Physics, University of Gothenburg, Origov\"agen 6B, 412 58 Gothenburg, Sweden
\\
\bf{8} Department of Chemistry, Stanford University, 333 Campus Drive, Stanford, CA 94305, USA.
\\
\bf{9} Department of Physics and Astronomy, Uppsala University, Uppsala, Sweden.
\\
\bf{10} Elettra-Sincrotrone Trieste, 34149 Basovizza, Trieste, Italy.
\\
\bf{11} Scuola Normale Superiore, I-56126 Pisa, Italy
\\
\bf{12} Institut für Physik und Astronomie, Universit\"at Potsdam, Karl-Liebknecht-Straße 24/25, DE-14476 Potsdam
\\
\bigskip
* thomas.wolf@stanford.edu, henrik.koch@sns.it, mguehr@uni-potsdam.de

\end{flushleft}

\section*{Abstract}
We present the first investigation of excited state dynamics by resonant Auger-Meitner spectroscopy (also known as resonant Auger spectroscopy) using the nucleobase thymine as an example. Thymine is photoexcited in the UV and probed with X-ray photon energies at and below the oxygen K-edge. After initial photoexcitation to a $\pi\pi$* excited state, thymine is known to undergo internal conversion to an $n\pi$* excited state with a strong resonance at the oxygen K-edge, red-shifted from the ground state $\pi$* resonances of thymine (see our previous study Wolf \textit{et al.}, \textit{Nat. Commun.}, 2017, \textbf{8}, 29). We resolve and compare the Auger-Meitner electron spectra associated both with the excited state and ground state resonances, and distinguish participator and spectator decay contributions. Furthermore, we observe simultaneously with the decay of the $n\pi$* state signatures the appearance of additional resonant Auger-Meitner contributions at photon energies between the $n\pi$* state and the ground state resonances. We assign these contributions to population transfer from the $n\pi$* state to a $\pi\pi$* triplet state via intersystem crossing on the picosecond timescale based on simulations of the X-ray absorption spectra in the vibrationally hot triplet state. Moreover, we identify signatures from the initially excited $\pi\pi$* singlet state which we have not observed in our previous study.


\section{Introduction}
\begin{figure}[h]
	\includegraphics[width=\textwidth]{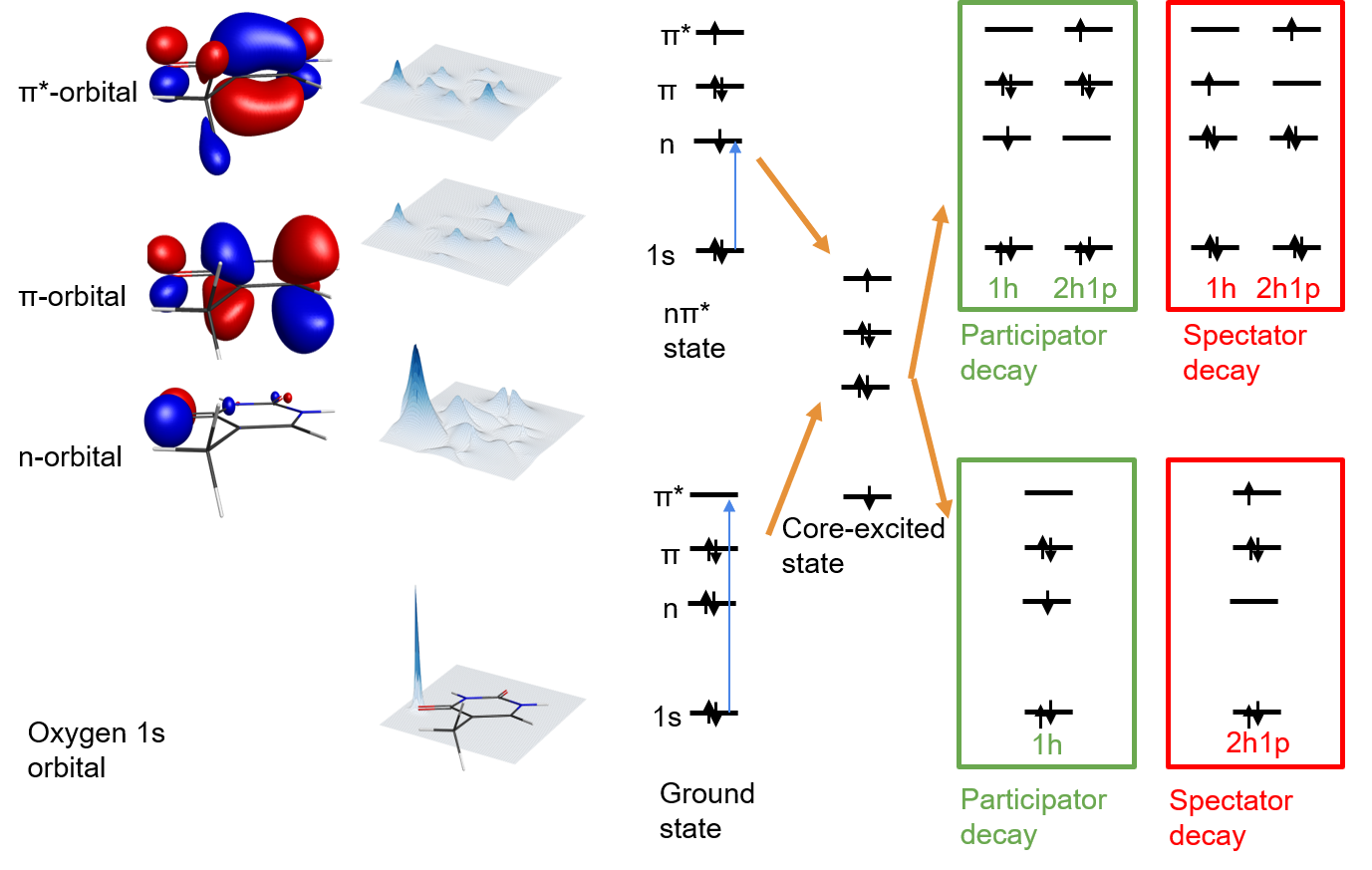}
	\caption{left: Highest occupied and lowest unoccupied molecular orbitals and oxygen 1s orbital of thymine and their electron density projections onto the thymine ring plane. Right: Participator and spectator decay channels in the 1-electron-picture starting from the electronic ground state and the $n\pi$* state. The core-excited state associated with the $n\pi$* resonant Auger-Meitner features is identical with the core-excited state of the ground state $\pi$* resonance associated with the oxygen(8) position (see Fig.~\ref{fig:RAM}). Participator decay channels from the ground state have 1-hole (1h) character, spectator decay channels 2-hole-1-particle (2h1p) character with respect to the initial state. The same applies to the the character of the final states from participator and spectator decay with respect to the initial $n\pi$* state. However, both participator and spectator decay from the $n\pi$* state can have 1h and 2h1p character with respect to the ground state electron configuration.}
	\label{fig:Scheme}
\end{figure}
Time-resolved soft X-ray spectroscopy is emerging as a powerful tool for the investigation of ultrafast dynamics in organic molecules~\cite{Attar2017,Pertot2017,Wolf2017,Wolf20172,Mcfarland2014,Neville2018,Bhattacherjee2017,Brausse2018,Inhester2019}. Energy separations of more than 100 eV in the K-edges of the most abundant elements in organic chemistry (excluding hydrogen), carbon, oxygen, and nitrogen allow photoinduced dynamics to be interrogated element and site-specifically~\cite{Wolf20192}.

We have recently investigated the ultrafast UV-induced dynamics of the nucleobase thymine using near-edge X-ray absorption fine structure (NEXAFS) spectroscopy at the oxygen edge~\cite{Wolf2017}. Thymine can be promoted to an excited electronic state, which is well characterized by a single electron excitation from its $\pi$ orbital to its $\pi$* orbital ($\pi\pi$* character, see Fig.~\ref{fig:Scheme}). It exhibits an additional low-lying and spectroscopically dark excited state characterized by single electron excitation from an oxygen lone pair (n) orbital to the $\pi$* orbital ($n\pi$* character, see Fig.~\ref{fig:Scheme}). These two low-lying excited states are connected by a $\pi\pi$*/$n\pi$* conical intersection. The existence and time scale of a relaxation channel by internal conversion through the $\pi\pi$*/$n\pi$* conical intersection was, however, under debate~\cite{Ullrich2004,Yu2016,Mai2017,Hudock2007,Barbatti2010,Yamazaki2012,Nakayama2013,Picconi2011,Asturiol2010}. By probing the excited state dynamics via \mbox{NEXAFS} spectroscopy at the oxygen K-edge, we were able to observe an unambiguous signature of the population of the $n\pi$* state within 60 fs after photoexcitation~\cite{Wolf2017}.

\begin{figure*}[t]
	\includegraphics[width=\textwidth]{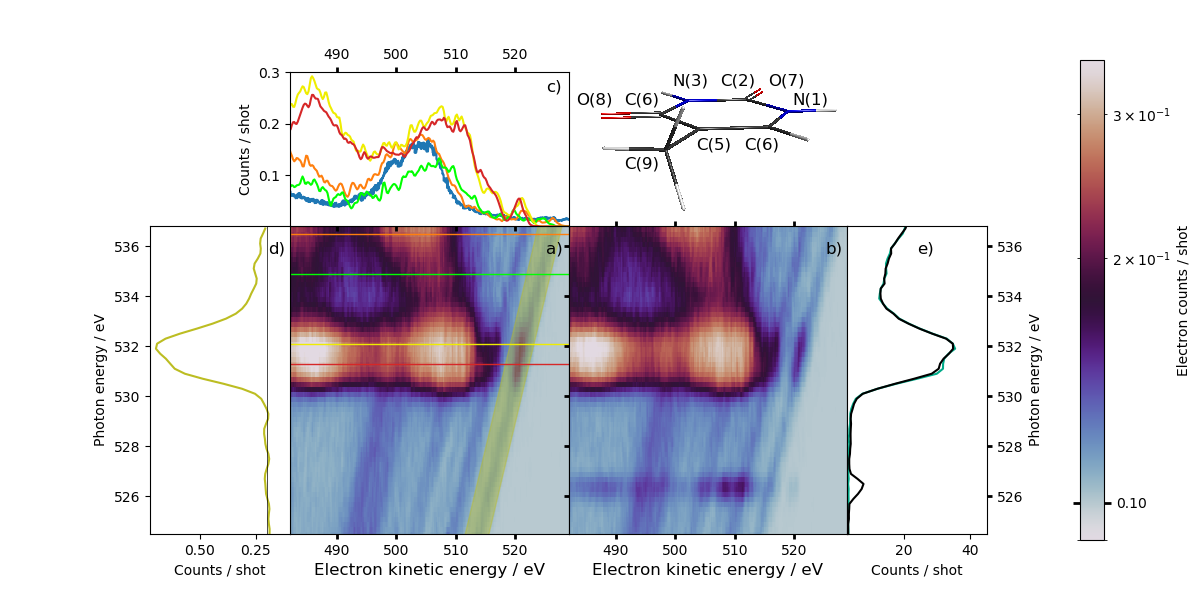}
	\caption{a) Steady-state resonant Auger-Meitner map of thymine. Diagonal lines refer to valence photoelectron signatures and participator Auger-Meitner decay channels (note the logarithmic color scale). b) Resonant Auger-Meitner map of thymine 4 ps after optical excitation ($\sim$13 \% excitation fraction). A pump-induced Auger-Meitner signature is visible at a photon energy of 526.5 eV. c) Auger-Meitner spectra at specific photon energies (marked by horizontal lines in part a)). For comparison, also the non-resonant Auger-Meitner spectrum from Ref.~\citen{Mcfarland2014} is included (blue). d) Photon energy-dependent, integrated diagonal area as marked in part (a). e) Near-edge X-ray Absorption Fine Structure (NEXAFS) spectra resulting from integrating the Auger-Meitner maps in part a) (light green) and b) (dark green) over all electron kinetic energies.  }
	\label{fig:RAM}
\end{figure*}

NEXAFS spectroscopy probes resonant transitions of core electrons to empty valence orbitals just below an elemental edge. The cross-sections of these resonances are strongly dependent on the overlap between the involved core and empty valence orbitals. Since the core orbitals are strongly localized (see the electron density projections in Fig.~\ref{fig:Scheme}), the cross-section is a sensitive probe for the localization of a valence orbital at the specific atomic site. The oxygen K-edge NEXAFS spectrum of thymine in its electronic ground state (see Fig.~\ref{fig:RAM} d)) exhibits a double-peak structure, where each of the two peaks can be associated with a specific oxygen site in the molecule~\cite{Plekan2008}. If thymine is photoexcited to the $\pi\pi$* state by a femtosecond UV laser pulse prior to the NEXAFS probe, the molecule exhibits an electron vacancy in the formerly fully occupied $\pi$ orbital. The vacancy opens a new NEXAFS resonance from the oxygen 1s orbitals. However, due to the strongly delocalized character of the $\pi$ orbital (see Fig.~\ref{fig:Scheme}), the corresponding NEXAFS transition was too weak to be observed in our experiment. With internal conversion through the $\pi\pi$*/$n\pi$* conical intersection and the corresponding electronic character change, the electron vacancy switches from the $\pi$ orbital to the oxygen lone pair (n) orbital.  The corresponding oxygen edge NEXAFS resonance is significantly stronger due to strong localization of the n-orbital at the oxygen site (see Fig.~\ref{fig:Scheme}). 

We obtained transient NEXAFS spectra of thymine by scanning the oxygen edge using a narrow-bandwidth X-ray Free Electron Laser (XFEL) source and integrating over the X-ray induced photoemission intensity, which is proportional to the X-ray absorbance in the sample. The photoemission kinetic energies of $>$480 eV contain electrons from valence photoionization, resonant Auger-Meitner (RAM) decay\footnote[1]{Auger-Meitner decay is traditionally known as Auger decay. Following the proposal from Ref.~\citen{Matsakis2019}, we are highlighting Lise Meitner's contributions to the discovery of this effect by using the term Auger-Meitner decay.}, and non-resonant Auger-Meitner decay. In our previous publication~\cite{Wolf2017}, we focused on the NEXAFS spectra resulting from integration over all electron kinetic energies. In the present report, we are investigating the underlying four-dimensional dataset (intensity vs. photon energy, electron kinetic energy, and pump-probe delay) in more detail to disentangle and better understand the individual contributions. The most prominent contributions to the  photoemission originate from RAM decay of core-excited molecules. 

A number of molecules have been investigated by RAM spectroscopy so far (see e.g. Ref.~\citen{Feifel2011} and Refs.~cited therein). Additionally, RAM spectroscopy has been used to study electron transfer on surfaces~\cite{Fohlisch2005}. Our report marks the first investigation of ultrafast photoinduced dynamics with RAM spectroscopy. RAM decay can be described as a resonant Raman scattering process~\cite{Gelmukhanov1999}: an Auger-Meitner electron is emitted from a molecule as a result of absorption of an X-ray photon at a NEXAFS resonance. It is, therefore, closely related to resonant inelastic X-ray scattering (RIXS)~\cite{Gelmukhanov1994,Platzman1998}. In contrast to RIXS, emission channels are not restricted by optical selection rules, which makes RAM spectroscopy potentially more versatile, but also more difficult to interpret. However, RIXS spectra of organic gas phase species are extremely difficult to obtain, since X-ray fluorescence cross-sections for light elements are small and existing X-ray spectrometers collect fluorescence only in a small solid angle which strongly limits the obtainable signal levels. In contrast, RAM decay is the dominant relaxation channel after soft X-ray absorption for the lighter elements of the periodic table, and RAM electrons can be detected with high efficiency.

In the following, we will present two-dimensional maps of photon energy-dependent RAM spectra of thymine in the electronic ground state, discuss the changes in the RAM spectra due to valence photoexcitation, and investigate their time-dependence. 

\section{Methods}

\subsection{Experimental methods}

The experimental procedure is described in detail in Ref.~\citen{Wolf2017}. In short, time-resolved NEXAFS spectroscopy was performed at the Linac Coherent Light Source (LCLS) free electron laser (FEL) facility, SLAC National Accelerator Laboratory, using the soft X-ray (SXR) instrument~\cite{Dakovski2015,Emma2010}. Thymine molecules were excited by ultrashort 267 nm pulses ($<$100 fs) derived from a Ti:Sa laser system via third harmonic generation. The laser system was synchronized to the FEL. Residual arrival time jitter between laser and FEL pulses is corrected via a single-shot cross-correlator~\cite{Beye2012}. The dependence of the excited state NEXAFS spectrum on the pump pulse fluence was scanned to ensure excitation in the linear regime. Soft X-ray pulses were used to probe the sample in the oxygen K-edge spectral region from 520 to 550 eV by simultaneously tuning the FEL and the monochromator of the SXR instrument with an energy resolution of $<$0.5 eV~\cite{Heimann2011}. Oxygen-edge photoemission was recorded with the 2 m long LCLS-FELCO (LCLS-FEL correlation) magnetic bottle spectrometer~\cite{Frasinski2013}. Soft X-ray pulses were delayed with respect to the UV pulses between $-$200 fs and 20 ps. To achieve NEXAFS difference spectra in the presence of strong fluctuations of the LCLS pulses in intensity and relative arrival time, the UV laser pulses were blocked on some shots, and data were recorded on a shot-by-shot basis for later resorting during the analysis.

\subsection{Theoretical methods}

While RAM spectra can be simulated for small molecules in their electronic ground state~\cite{Holzmeier2018}, there are, to our knowledge, no methods available for simulating excited state RAM spectra for medium sized organic molecules like thymine. Thus, we focus on the evaluation of relative energies of RAM decay final states and the simulation of hot ground state and triplet state NEXAFS signatures. The nine lowest cationic states of Thymine were calculated using CC3/aug-cc-pVDZ~\cite{eTcode,Paul2020,Kendall1992}. Among those states, two ionization vectors were dominated by 2h1p-contributions. Using the CCSD density of the final state with the lower energy, 
the state could be characterized as an $n\pi\pi^\star$ state.

Simulations of the NEXAFS spectra of the cold ground state, the hot ground state, and the hot triplet state of thymine are performed similarly to the simulation of valence photoelectron spectra in Ref.~\citen{Wolf2019} using the same 50 random geometries per simulated spectrum as in Ref.~\citen{Wolf2019}, which were sampled from ab initio molecular dynamics simulations. Transition energies and oscillator strengths were calculated at CCSD/aug-cc-pVDZ level\cite{CVS:jcp:2015,Faber:Xray:triplet,eTcode}.

\section{Results and discussion}

\subsection{Ground state resonant Auger-Meitner decay}

Figure \ref{fig:RAM} a) shows a 2D false-color plot of electron intensity vs. electron kinetic energy and  X-ray photon energy from thymine molecules in their electronic ground state. The plot contains two types of features: intense signatures from resonant and non-resonant Auger-Meitner decay with non-trivial photon energy dependence, and weak signatures, which shift linearly with the photon energy in kinetic energy. Integration of the 2D plot over all electron kinetic energies yields the ground state oxygen edge NEXAFS spectrum as in Refs.~\citen{Wolf2017,Plekan2008} (see light green spectrum in Fig.~\ref{fig:RAM} e)). The spectrum shows a double peak structure associated with resonant oxygen 1s$\xrightarrow{}\pi$* excitations from the oxygen(8) position (531.4 eV) and oxygen(7) position (532.2 eV, see Fig.~\ref{fig:RAM} for nomenclature).

\begin{figure*}[t]
	\includegraphics[width=\textwidth]{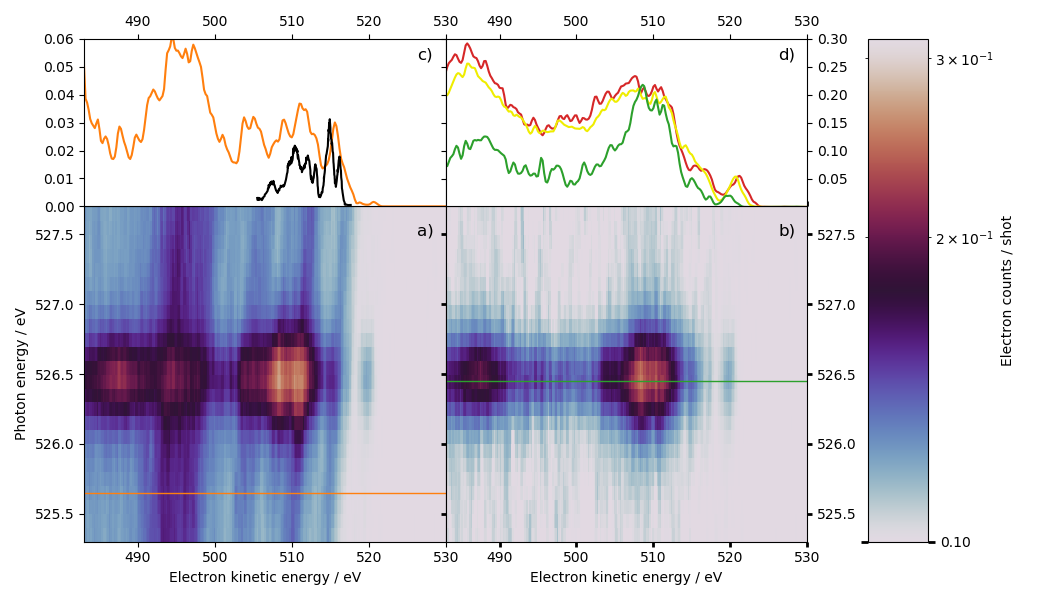}
	\caption{a) Detailed Resonant Auger-Meitner map of thymine in the vicinity of the UV-induced $n\pi$* state resonance, 1.2 ps after optical excitation. A weak signature at electron kinetic energies above the ground state photolines is visible at the resonance photon energy. b) Same as part a), but after subtraction of a steady-state Auger-Meitner map. c) Comparison of the photolines (see line in part a)) with a valence photoelectron spectrum from Ref.~\citen{Trofimov2006}, which is shifted to account for the difference in ionization photon energy. d) Comparison of the Auger-Meitner signatures at the UV-induced resonance with the $\pi$* resonance Auger-Meitner signatures (see Fig.~\ref{fig:RAM}). }
	\label{fig:Ex_RAM}
\end{figure*}

The weak signatures with linear photon energy dependence are photolines from valence electron ionization of thymine. A comparison of the photolines at a photon energy of 525 eV (red) with a high-resolution valence photoelectron spectrum adapted from Ref.~\citen{Trofimov2006} (black) is shown in Fig.~\ref{fig:Ex_RAM} c). The latter is linearly shifted to account for the different ionization photon energy (80 eV). The region between 505 eV and 517 eV electron kinetic energy (20 eV to 8 eV electron binding energy) in our spectrum agrees well with the literature spectrum taking into account the different photon energies used for the valence ionization. The retardation voltage applied to our spectrometer (480 V) was optimized for collection of a large energy range of the oxygen-edge RAM spectra. Therefore, our energy resolution for the most loosely bound  valence photoelectrons (35 eV kinetic energy after retardation) is limited. Thus, e.g. the two peaks with lowest binding energy (514 and 515 eV electron kinetic energy) in the literature spectrum corresponding to ionization from the $\pi$ and the n orbital cannot be fully resolved in our spectra and instead appear as a peak with a shoulder.

The photoelectron spectrum of thymine up to 20 eV binding energy (505 eV kinetic energy in Fig.~\ref{fig:Ex_RAM} c))  has been investigated by a number of studies~\cite{Trofimov2006,Dougherty1976,Schwell2014,Fulfer2015}. It is known that the Koopmans picture of ionization from a single molecular orbital typically starts to break down for hydrocarbons at around 20 eV electron binding energy~\cite{Trofimov2006,Cederbaum1978}. The region between 20 eV and 40 eV binding energy (505 eV to 485 eV electron kinetic energy in Fig.~\ref{fig:Ex_RAM} c), which has not been investigated previously, is therefore difficult to interpret without high-level simulations. It can be expected to exhibit contributions from shake-up ionization channels and double ionization. Nevertheless, the onset of the 30 eV binding energy peak (495 eV in Fig.~\ref{fig:Ex_RAM} c)) towards higher binding energies (40 eV binding energy, 485 eV electron kinetic energy) coincides with the Koopmans IPs of the innermost valence orbitals. 

The intense and non-dispersing signatures in Fig.~\ref{fig:RAM} a), which obviously contribute the majority of the intensity to the ground state NEXAFS spectrum (Fig.~\ref{fig:RAM} e)) are due to Auger-Meitner decay processes. The map shows strong horizontal features at the photon energies of the split oxygen pre-edge $\pi$* resonance at photon energies of 531.4 eV and 532.2 eV with two broad and intense maxima centered at 507 eV and 485 eV electron kinetic energy (see yellow and red spectra in Fig.~\ref{fig:RAM} c). 

It exhibits an intensity minimum at a photon energy of $\approx$534 eV and some features in a photon energy region between 534 eV and the onset of the oxygen ionisation continuum in thymine at 537.3 eV (not shown in Fig.~\ref{fig:RAM} a)), which coincide with additional resonances to valence orbitals with strong Rydberg character~\cite{Plekan2008}. At photon energies approaching the ionization edge, additional features with increasing intensity and a broad maximum at 504 eV are observable.

The Auger-Meitner decay features can be categorized into signatures readily below the oxygen K-edge ($\pi$* resonances), which originate from the RAM decay and signatures at the oxygen K-edge. The latter belong to a transition regime of excitation into high-lying Rydberg orbitals, orbitals above the ionization threshold, and threshold 1s photoionization. The shift of the maximum with the highest electron kinetic energy from 507 eV ($\pi$* resonances) below the K-edge to 504 eV above the K-edge (see Fig.~\ref{fig:RAM} c)) clearly demonstrates the transition from the combination of core-excited intermediate state and valence-ionized final state of the RAM decay to the combination of core-ionized intermediate state and valence doubly-ionized final state of non-resonant Auger-Meitner decay. Electrons from non-resonant Auger-Meitner decay with doubly charged final states must overcome a substantially larger Coulomb potential than electrons from RAM decay, where the final state is singly charged. Additionally, the high kinetic energy peak of the Auger electron spectra at and above the K-edge almost perfectly coincides with the global maximum of the non-resonant Auger spectrum measured at 565 eV photon energy (see blue spectrum in Fig.~\ref{fig:RAM} c) taken from Ref.~\citen{Mcfarland2014}. There was an error in the calibration of the spectrum published in Ref.~\citen{Mcfarland2014} as pointed out in Ref.~\citen{Wolf20172}).

RAM decay channels and final states from the electronic ground state can be divided into participator and spectator decay channels~\cite{Langer1997}. Participator decay involves the electron which participated in the core-excitation and an additional valence electron in the core hole filling and electron emission. The electrons involved in spectator decay are both different from the electron which participated in the core-excitation (see Fig.~\ref{fig:Scheme}). Thus, the electron configuration of the participator decay final state only differs by a valence electron hole from the initial (ground) state prior to X-ray absorption (1-hole state, see Fig.~\ref{fig:Scheme}). They are identical to final states from valence photoionization. In contrast, the electron configurations of spectator decay final states differ from that of the initial state by at least two valence electron holes and the core-excited electron (2-hole-1-particle states).

Since the participator decay features exhibit identical final states with valence ionization, they are easily identified as intensity modulations of the photolines~\cite{Carravetta1997}. Their linear shift in kinetic energy with photon energy (dispersion) is a well-known effect arising from the Raman-scattering character of the RAM decay process~\cite{Gelmukhanov1999}. The majority of the features at 507 eV and 485 eV are clearly not due to intensity modulation of photolines in this electron kinetic energy regime. Therefore, they can be assigned to spectator decay channels.

The photon energy-dependent intensity of the photoline with highest kinetic energy is plotted in Fig.~\ref{fig:RAM} d). The intensity modulation by the participator decay channels in the photon energy range of the $\pi$* resonances is clearly visible. At photon energies above the $\pi$* resonances, there is still a slight enhancement in the intensities of both photolines of Fig.~\ref{fig:RAM} d). This is a hint for RAM enhancement through the aforementioned Rydberg resonances closer to the oxygen K-edge.

\subsection{Excited state resonant Auger-Meitner spectra}
\begin{figure*}[t]
	\includegraphics[width=\textwidth]{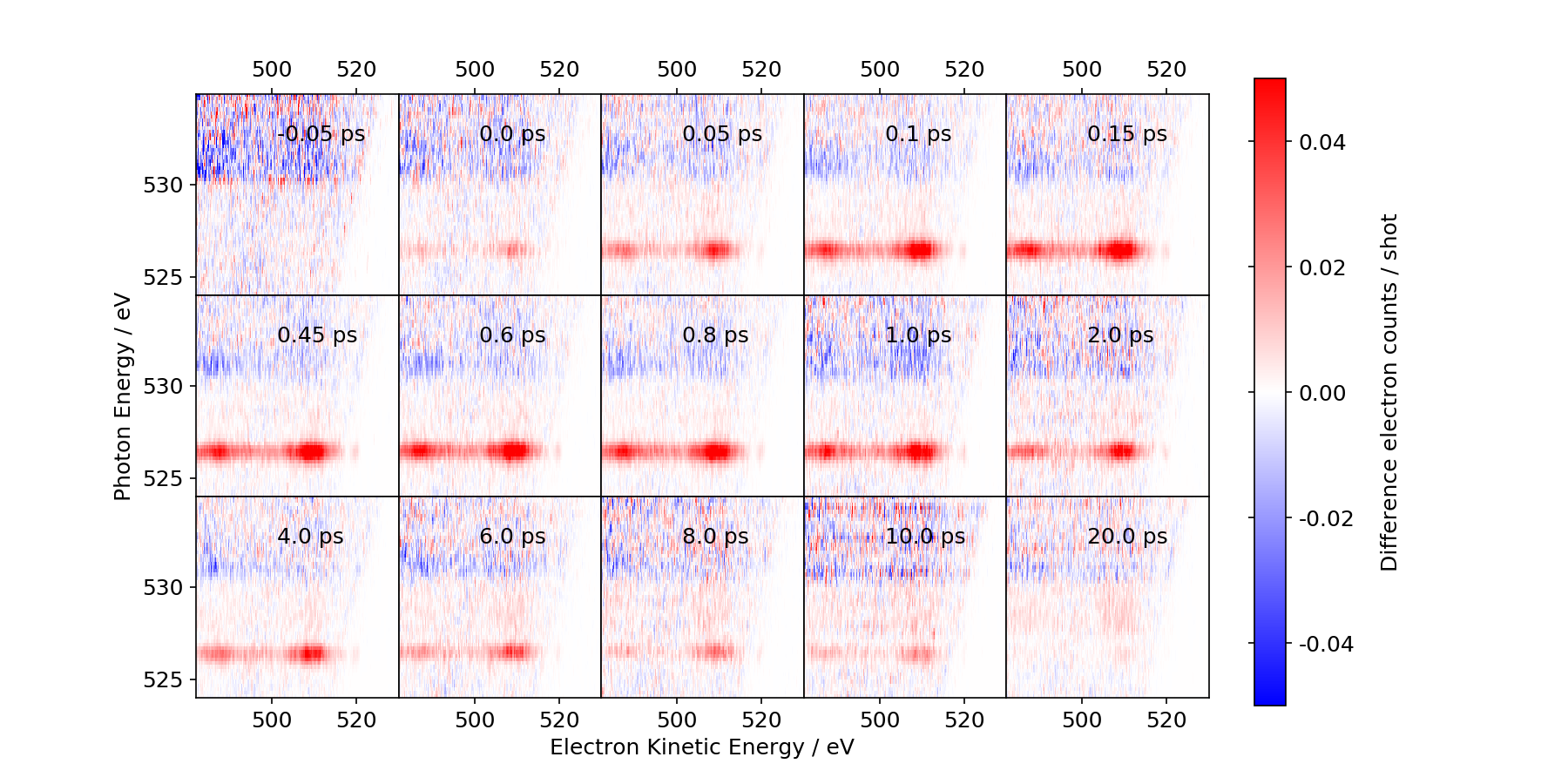}
	\caption{2D difference Auger-Meitner maps for different pump-probe delays. Around time zero, a negative signature in the photon energy regime of the ground state $\pi$* resonances is visible. The positive $n\pi$* state signature appears with a slight delay, which is connected to the $\pi\pi$*/$n\pi$* internal conversion time scale. The intensity of the $n\pi$* Auger-Meitner signature decays on the picosecond timescale. Simultaneously, a positive signature with weak photon energy dependence and a double peak structure in kinetic energy becomes more intense at photon energies in between the ground state $\pi$* resonances and the $n\pi$* state signature.}
	\label{fig:Diff_RAM}
\end{figure*}
If the RAM decay is initiated from a valence excited state, the additional electron configuration change from the photoexcitation (in the single electron picture) is influencing the electron configurations of the RAM final states. Again, the final states of participator decay have 1-hole character, the spectator decay states have 2-hole-1-particle character with respect to the initial state of the RAM decay. Since the initial state is now a valence excited state, both the participator and spectator decay final states can have 1-hole or 2-hole-1-particle character with respect to the electronic ground state, dependent on the participation of the valence-excited electron in the decay (see Fig.~\ref{fig:Scheme}). 

Figure \ref{fig:RAM} b) shows a photoemission map analogous to Fig.~\ref{fig:RAM} a), but recorded 4 ps after ~13 \% of the thymine molecules were valence-excited with a femtosecond UV pulse (for details see Refs.~\citen{Wolf2017,Wolf2019}). The valence excitation yields an obvious new feature in the photoemission map, a signature at 526.5 eV photon energy, shifted from the ground state $\pi$* resonances by roughly the energy of the absorbed UV photon (4.65 eV). Summation over all electron kinetic energies yields the photoexcited NEXAFS spectrum from Ref.~\citen{Wolf2017} (see the dark green spectrum in Fig.~\ref{fig:RAM} e)). Comparison of the ground state and the photoexcited spectra in Fig.~\ref{fig:RAM} e) reveals not only a substantial intensity increase at 526.5 eV, but also an additional change, a slight bleach in the lower photon energy ground state $\pi$* resonance, which is not obvious from the photoemission maps in Fig.~\ref{fig:RAM} a) and b). 

The signature at 526.5 eV obviously must be due to additional, UV-induced resonant Auger decay processes. As detailed in Ref.~\citen{Wolf2017}, it is due to resonant excitation of an oxygen 1s electron into the oxygen lone pair (n) orbital of thymine, which exhibits an electron hole in the valence-excited $n\pi$* state electron configuration. The core-excited intermediate state of this process is, thus, identical to the intermediate state of the $\pi$* resonance associated with the oxygen(8) position (see Figs.~\ref{fig:Scheme} and \ref{fig:RAM}). Taking into account that the 526.5 eV signature originates only from UV-excited 13 \% of the molecules, its cross-section can be estimated to be comparable to the ground state $\pi$* resonance cross-sections.

Figure \ref{fig:Ex_RAM} a) shows a more detailed photoemission map 1.2 ps after valence photoexcitation in the area of the $n\pi$* state signature at 526.5 eV. Figure \ref{fig:Ex_RAM} b) shows the same area, but after subtraction of a photoemission map without valence photoexcitation. The $n\pi$* state resonant Auger signature features two maxima at 509 eV and 487 eV. Based on their similarity in width and position with the spectator peaks of the ground state RAM spectra (see Fig.~\ref{fig:Ex_RAM} d)), we assign them to spectator decay channels. The differences in peak width, maxima and relative intensity with respect to the ground state spectator signatures requires simulation of ground and excited state spectator decay, which is beyond the scope of this work (see theoretical methods).

Apart from the intense $n\pi$* spectator decay signatures, no enhancement of photolines is observable in the photon energy regime of the $n\pi$* state resonance in Fig.~\ref{fig:RAM} and Fig.~\ref{fig:Ex_RAM} a). This is expected, since the photolines originate from valence ionization of the 87 \% molecules which are not photoexcited. The 13 \% photoexcited population should in principle yield an additional set of photolines with higher kinetic energies than the ground state features analog to excited-state  spectra in time-resolved valence photoelectron spectroscopy. However, these signatures are too weak to be observable at the signal-to-noise level of our data. We observe, however, a new feature at a kinetic energy of 520 eV, beyond the ground state photolines right at the $n\pi$* resonance. Its relative shift to higher kinetic energies by 2.4 eV with respect to the ground state photolines (see Fig.~\ref{fig:Ex_RAM} c)) fits very well to the $n\pi$* state signature observed in valence photoelectron spectroscopy~\cite{Wolf2019}. It can, therefore, be assigned to the same final state, the second lowest cationic state of thymine, which exhibits 1s$^{2}$n$^{1}$$\pi^{2}$$\pi^{*}{}^{0}$ configuration (see 1h final state of excited state participator decay in Fig.~\ref{fig:Scheme}). Thus, it is the signature of an $n\pi$* state participator decay channel involving the n electron from core excitation and the $\pi$* electron from valence excitation. Hence, the participator line represents the valence photoelectron spectrum modulated by the cross-section of the core-to-n resonance. As a consequence, RAM spectroscopy can be instrumental in assigning photoelectron spectroscopy signatures to specific electronic states.

The next lowest kinetic energy participator decay channel of the $n\pi$* resonance must involve the $\pi$ orbital (1s$^2$n$^1\pi^2\pi^*{}^1\xrightarrow{}$1s$^1$n$^2\pi^2\pi^*{}^1\xrightarrow{}$1s$^2$n$^1\pi^1\pi^*{}^1$). The corresponding cationic final state exhibits a 2-hole-1-particle configuration with respect to the thymine ground state. According to our calculations, its binding energy is 3.5 eV higher than the 1s$^{2}$n$^{1}$$\pi^{2}$$\pi^*{}^{0}$ final state, which is in good agreement with a shoulder in the higher kinetic energy spectator decay peak at 515 eV (see Fig.~\ref{fig:Ex_RAM} d)). 

\subsection{Time-dependence of Auger-Meitner decay signatures}

Figure \ref{fig:Diff_RAM} shows difference RAM maps for different pump-probe delays at photon energies covering the $\pi$* resonances as well as the $n\pi$* state signature. Around time zero, a negative signature appears in the photon energy regime of the ground state $\pi$* resonances. These signature originates from a bleach of the ground state NEXAFS spectrum due to the excitation of population to valence-excited states. With a slight delay (60 fs)~\cite{Wolf2017}, the $n\pi$* state signature appears in the difference RAM maps. Its intensity decays on the picosecond timescale in agreement with Ref.~\citen{Wolf2017}. Simultaneously, additional positive features develop at photon energies between the $\pi$* resonances and the $n\pi$* resonance. These features appear as an unspecific increase in the baseline of the difference-NEXAFS spectra of Ref.~\citen{Wolf2017}. However, when dispersed in electron kinetic energy, a double peak structure with maxima at comparable kinetic energies to the $n\pi$* feature can be identified (see Fig.~\ref{fig:Ex_RAM} and supplementary figure 1). Thus, the features are clearly due to RAM decay.
\begin{figure}[t]
	\includegraphics[width=\textwidth]{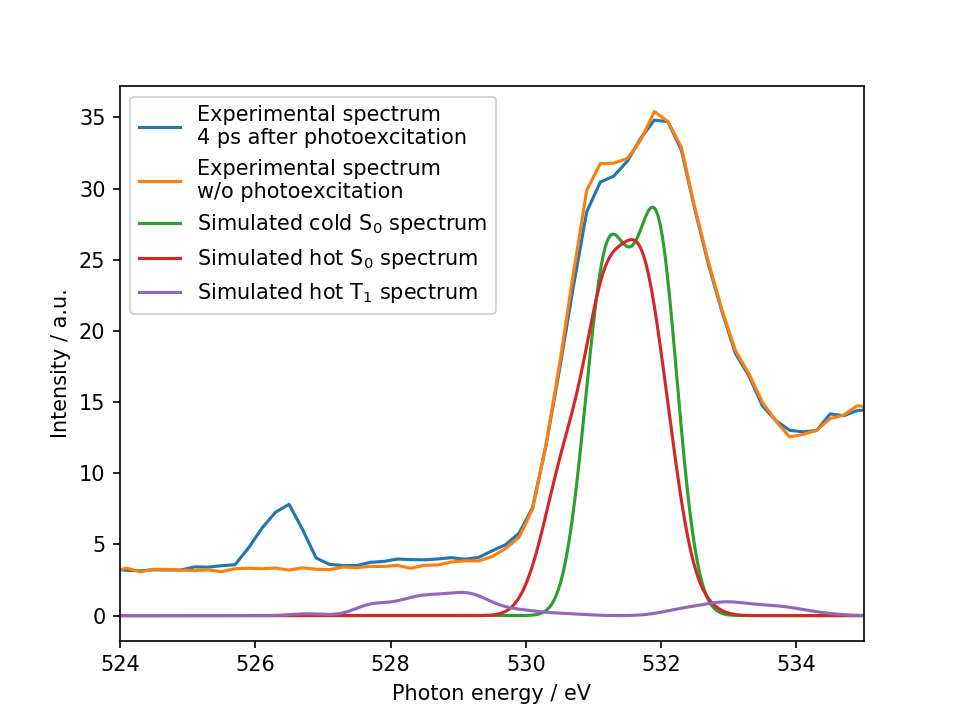}
	\caption{Comparison of the experimental spectra 4 ps after photoexcitation and withouth photoexcitation with simulated spectra of the hot and cold singlet ground state (S$_0$) and the hot triplet ground state (T$_1$). Simulated spectra are based on CCSD/auc-cc-pVDZ calculations of NEXAFS transitions. For details see the methods section.}
	\label{fig:Comp}
\end{figure}

The relative timing of their intensity increase with the decrease of the $n\pi$* resonance Auger-Meitner features suggests that they correspond to the species the $n\pi$* population decays to. In principle, decay from the $n\pi$* state can lead to two different states, the hot singlet ground state and the hot triplet $\pi\pi$* state. In our earlier valence photoelectron study, we could clearly observe population of the triplet $\pi\pi$* state~\cite{Wolf2019}. However, the existence of a relaxation channel to the singlet ground state cannot be excluded.

To give an unambiguous assignment of the experimental features to an electronic state, we performed simulations of the NEXAFS spectra of the hot ground state and the hot triplet state of thymine assuming a statistical distribution of the absorbed photon energy over all vibrational degrees of freedom (see methods). For comparison with the experimental NEXAFS spectra of thymine, we also simulated its room temperature spectrum. A comparison of simulated and experimental spectra is depicted in Fig.~\ref{fig:Comp}. 

The $\pi$* resonance double peak structure of the experimental ground state spectrum is well reproduced in the simulated room temperature spectrum. The $\pi$* resonance peak positions of the hot ground state spectra are very similar to the room temperature spectrum, but show a slight broadening. It is, however, very clear that the hot ground state spectrum cannot explain the broad and weak experimental signatures in between the $n\pi$* resonance at 526.5 eV and the $\pi$* resonances at 531.4 eV and 532.2 eV (see comparison of experimental spectra with and without photoexcitation in Fig.~\ref{fig:Comp}). In contrast, the simulated hot triplet state spectrum shows a broad peak at 528.7 eV, which covers approximately the photon energy region, where the experimental signatures were observed. We can therefore assign the appearance of the features in between the $n\pi$* and the $\pi$* resonance to population of the $\pi\pi$* triplet state of thymine. Nevertheless, we also cannot exclude here additional hot ground state population.
\begin{figure}[t!]
	\includegraphics[width=\textwidth]{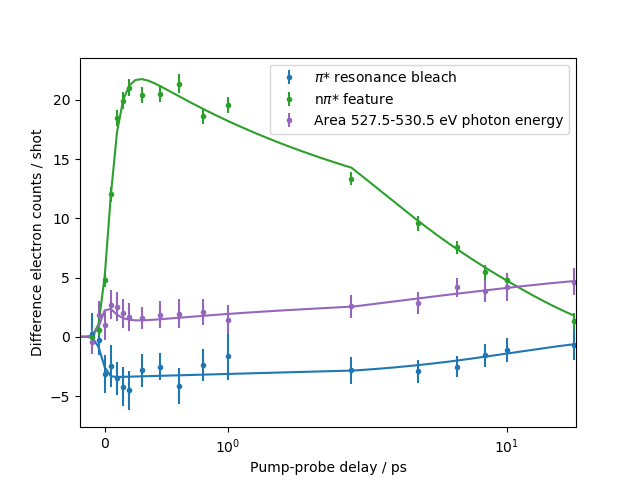}
	\caption{Time-dependent difference-photoemission yields in the photon energy regimes of the ground state $\pi$* resonance, the $n\pi$* state feature and the area in between together with fits using the same model as in Ref.~\citen{Wolf2017}. Note the logarithmic pump-probe delay axis.}
	\label{fig:Int}
\end{figure}

The time-dependent difference photoemission yields for the photon energy regimes of the $\pi$* resonance (bleach feature), the $n\pi$* feature, and the region in between are represented in Fig.~\ref{fig:Int} together with a fit based on the same model and time constants as in Ref.~\citen{Wolf2017}. The model assumes a consecutive decay from the $\pi\pi$* excited state to the $n\pi$* state with a time constant of 60 fs and further from the $n\pi$* state to the triplet state. The latter decay is biexponential with time constants of 1.9 ps and 10.5 ps. The intensity increase in the area between the $n\pi$* feature and the $\pi$* resonance on the picosecond timescale is well-described using the same two time constants as the intensity decay of the $n\pi$* feature, which serves as additional evidence for the assignment to population transfer from the $n\pi$* state to the $\pi\pi$* triplet state.

Additionally, the area in between the $\pi$* resonance and the $n\pi$* feature shows a sub-100 fs response at time zero, which is well described by the model used to extract the delayed rise of the $n\pi$* resonance in Ref.~\citen{Wolf2017} i.e. the amplitude rises simultaneously with the amplitude of the $\pi$* resonance bleach and decays simultaneously with the rise of the $n\pi$* feature. Due to its fast timescale, it is unlikely that these sub-100 fs modulations originate from population of the triplet state. Instead, the synchronization with the two other features strongly suggests it originating from population in the initially excited $\pi\pi$* state, which we have not directly observed so far in the NEXAFS spectra. According to our previous simulations, the lowest NEXAFS feature of the $\pi\pi$* state (1s$\xrightarrow{}\pi$) almost coincides with the NEXAFS feature of the $n\pi$* state (1s$\xrightarrow{}$n), but exhibits by a factor of 40 lower intensity. However, the $\pi\pi$* state must exhibit absorption signatures to higher-lying core-excited states in the photon energy regime between the $n\pi$* feature and the $\pi$* bleach. Our previous calculations of excited state NEXAFS spectra did not reveal features in this spectral region~\cite{Wolf2017}.
However, while the CCSD theory level of our previous calculations is well equipped to accurately describe the 1s$\xrightarrow{}$n and 1s$\xrightarrow{}\pi$ transitions, it can miss transitions to higher core-excited states, as discussed e.g.~in Refs.~\citen{TR:XAS:Pyrazine,List2020}.

\section{Conclusions}

In conclusion, we present the first investigation of resonant Auger-Meitner decay as an observable to study ultrafast excited state dynamics in organic molecules. The excited state spectator decay features are broad and do not show distinctive differences with respect to the ground state signatures. Their interpretation would require comparison with accurate simulations, which require significant method development. In contrast, the final states from the participator decay appear as relatively sharp lines which can be directly related to complementary valence photoelectron spectroscopy experiments. The information content of future time-resolved RAM spectroscopy investigations can be significantly increased by improving the kinetic energy resolution of the participator decay channels.

At large pump-probe delays, we observe clear Auger-Meitner decay signatures which are not obvious from the already published NEXAFS results. Based on simulations we assign them to the relaxation of thymine to the $\pi\pi$* triplet state in agreement with earlier results from time-resolved photoelectron spectroscopy. Furthermore, we observe a sub-100 fs response which we assign to population in the initially excited $\pi\pi$* singlet state, which we have not directly observed in the NEXAFS spectra before. 


\section*{Acknowledgments}
We thank Andrea Battistoni, Christoph Bostedt, Kelly Gaffney, Jakob Grilj,  Adi Natan, Timur Osipov, and Rob M. Parrish for contributions to the thymine data and useful discussions. This work was supported by the U.S. Department of Energy, Office of Science, Basic Energy Sciences, Chemical Sciences, Geosciences, and Biosciences Division, Atomic, Molecular, and Optical Science Program. Parts of this research were carried out at the Linac Coherent Light Source (LCLS) at the SLAC National Accelerator Laboratory. LCLS is an Office of Science User Facility operated for the U.S. Department of Energy Office of Science by Stanford University. M.G. acknowledges funding via the Office of Science Early Career Research Program through the Office of Basic Energy Sciences, U.S. Department of Energy and NB under Grant No. DE-SC0012376. M.G. is now funded by a Lichtenberg Professorship from the Volkswagen foundation. T.J.A.W. thanks the German National Academy of Sciences Leopoldina for a fellowship (Grant No. LPDS2013-14). R.F. and R.J.S. thank the Swedish Research Council (VR) and the Knut and Alice Wallenberg Foundation, Sweden for financial support. A.C.P., H.K., and S.C. acknowledge support from the Marie Sk{\l}odowska-Curie
European Training Network \emph{COSINE-COmputational Spectroscopy In Natural sciences and Engineering} (Grant
Agreement 765739). A.C.P., H.K., S.D.F., S.C. and R.H.M acknowledge the Research Council of Norway through FRINATEK Projects 263110 and 275506. S.C. acknowledges support from the Independent Research Fund Denmark (DFF-RP2 Grant 7014-00258B).


\bibliography{library}

\bibliographystyle{rsc}

\end{document}